\documentclass[useAMS,usenatbib]{mn2e} \voffset=-1cm

\usepackage{times,graphicx,aas_macros,color,subfigure,amssymb}

\newcommand{\myr}{${\rm M_{\sun}\,yr^{-1}}$}

\newcommand{\um}{$\umu$m}

\begin{document}

\title[A submm survey of RCS\,2319]{A submillimetre-bright $\mathbf{z\sim3}$ overdensity behind a $\mathbf{z\sim1}$ supercluster revealed by SCUBA--2 and \textit{Herschel}\thanks{Herschel is an ESA space observatory with science instruments provided by European-led Principal Investigator consortia with important participation from NASA. }}

\author[Noble et al.]{\parbox[h]{\textwidth}{A.\ G.\ Noble$^1$\thanks{E-mail:
nobleal@physics.mcgill.ca}, J.\ E.\ Geach$^{1,2}$, A.\ J.\ van Engelen$^3$,
T.\ M.\ A.\ Webb$^1$, K.\ E.\ K.\ Coppin$^{1,2}$, A.~Delahaye$^1$, D.\ G.\
Gilbank$^4$, M.\ D.\ Gladders$^5$, R.~J.~Ivison$^{6,7}$, Y.\ Omori$^1$, H.\
K.\ C.\ Yee$^8$} \vspace*{6pt}\noindent\\ $^1$Department of Physics, McGill
University, 3600 Rue University, Montr\'eal, Qu\'ebec, H3A\ 2T8, Canada \\
$^2$Centre for Astrophysics Research, Science \& Technology Research
Institute, University of Hertfordshire, Hatfield, AL10 9AB\\ $^3$Department of
Physics and Astronomy, Stony Brook University, Stony Brook, NY, 11794-3800,
USA\\ $^4$South African Astronomical Observatory, PO Box 9, Observatory, 7935,
South Africa\\ $^5$Department of Astronomy and Astrophysics, University of
Chicago, 5640 S. Ellis Ave., Chicago, IL, 60637, USA\\ $^6$UK Astronomy
Technology Centre, Science and Technology Facilities Council, Royal
Observatory, Blackford Hill Edinburgh, EH9 3HJ, UK\\ $^7$Institute for
Astronomy, University of Edinburgh, Blackford Hill Edinburgh, EH9 3HJ, UK\\
$^8$Department of Astronomy and Astrophysics, University of Toronto, 50 St
George Street, Toronto, Ontario M5S 3H4, Canada}

\date{}

\pagerange{\pageref{firstpage}--\pageref{lastpage}}\pubyear{2013}

\maketitle

\label{firstpage}

\begin{abstract} We present a wide-field (30$'$ diameter) 850\um\ SCUBA-2
map of the spectacular three-component merging supercluster, RCS\,231953+00,
at $z=0.9$. The brightest submillimetre galaxy (SMG) in the field ($S_{\rm
850}\approx12$\,mJy) is within 30$''$ of one of the cluster cores (RCS\,2319--C),
and is likely to be a more distant, lensed galaxy. Interestingly, the wider
field around RCS\,2319-C reveals a local overdensity of SMGs, exceeding the
average source density by a factor of 4.5, with a $<1$\% chance of being found in a random field. Utilizing \textit{Herschel}-SPIRE
observations, we find three of these SMGs have similar submillimetre colours.  We fit their observed 250--850\um\ spectral energy distributions to
estimate their redshift, yielding $2.5<z<3.5$, and calculate prodigious star formation rates (SFRs) ranging from $500-2500$\,\myr.  We speculate that these galaxies are either lensed SMGs, or signpost a physical structure at $z\approx3$: a `protocluster' inhabited by
young galaxies in a rapid phase of growth, destined to form the core of a
massive galaxy cluster by $z=0$.

\end{abstract} \begin{keywords}galaxies: clusters: individual (RCS\,231946+0030.6) -- galaxies: formation -- galaxies: evolution -- galaxies: high-redshift -- submillimetre: galaxies \end{keywords}

\section{Introduction}

Submillimetre (submm) surveys have a history of exciting revelations, beginning with the discovery of a population of submm-bright galaxies (SMGs) over
a decade ago \citep{Smail97, Barger98, Hughes98}. These SMGs are now known to
be high-$z$ \citep{Chapman05, Wardlow11}, gas-rich \citep{Frayer98, Frayer99} systems undergoing intense episodes of star formation, and
are the likely progenitors of massive elliptical galaxies seen locally
\citep{Eales99, Lilly99}.  The latest generation of submm telescopes, namely the \textit{Herschel Space Observatory} \citep{Pilbratt10} and SCUBA-2 \citep{Holland13} on the James Clerk Maxwell Telescope (JCMT), are ushering in a new era of submm astronomy, yielding samples of thousands of SMGs  \citep{Eales10HATLAS, Oliver12} and promising many discoveries  \citep{Chen13, Geach13, Casey13}.

The wide-field mapping power of SCUBA-2 opens up a new parameter space in
submm studies: the capability to survey volumes that sample the full range of
galaxy environment at high redshifts. The densest regions---the nodes in the
cosmic web---at all epochs are rare. However, it is essential to understand
the evolution of galaxies within these environments if we are to link the growth of galaxies with the large scale structure they inhabit. Submm studies of distant `protoclusters' are particularly promising since it is
becoming clear that the massive tail of the galaxy cluster `red-sequence'
assembles quickly at $z>2$ \citep{Papovich10}, possibly via intense starbursts
within gas-rich galaxies located in overdense (but potentially
pre-virialised) structures at high-$z$ \citep{Daddi09}. This
activity might well be heavily dust-obscured, requiring submm and/or infrared (IR) observations to detect.

\begin{figure*} \hspace{-105mm}
\subfigure{\includegraphics[width=0.6\textwidth]{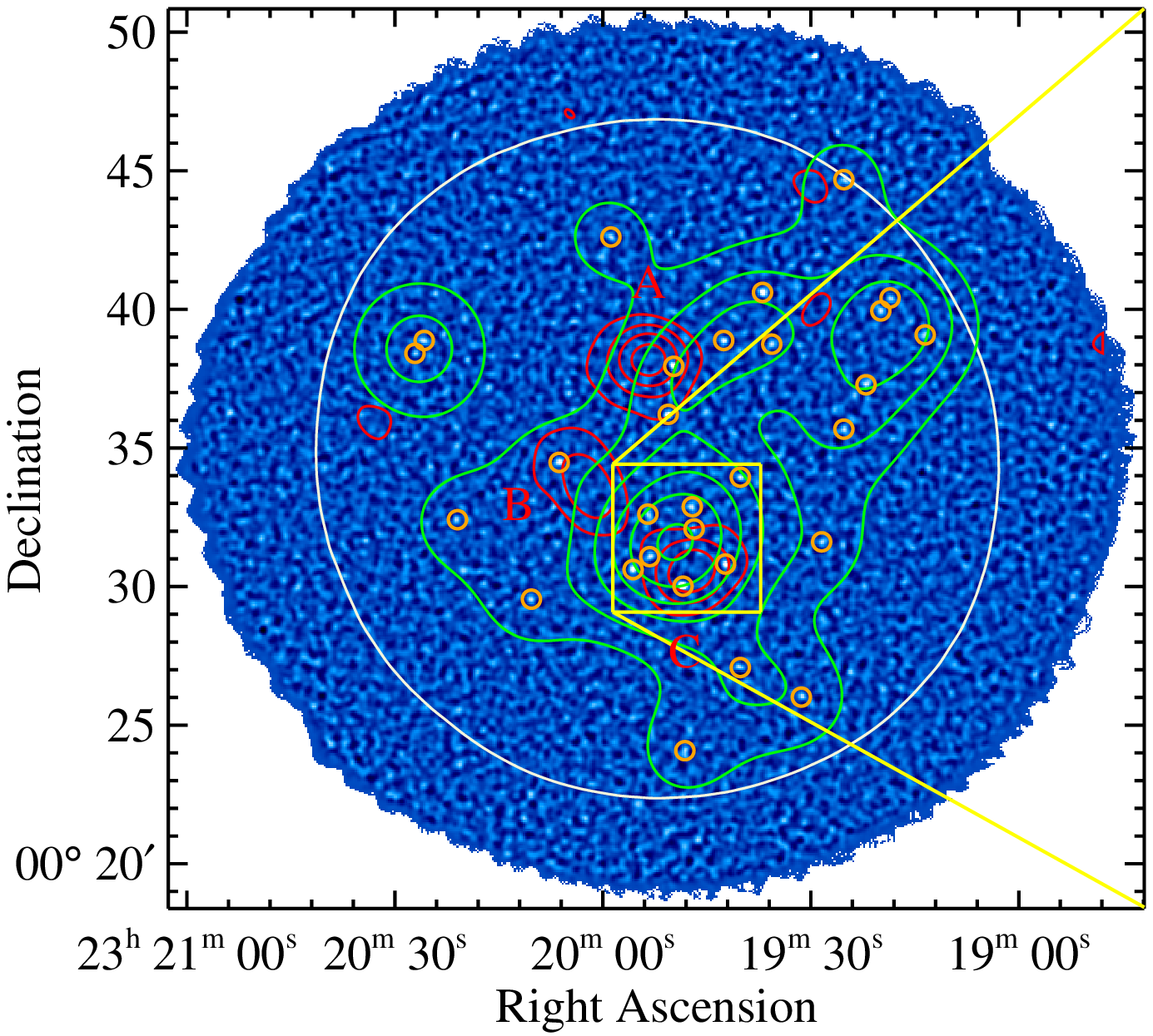}
}\hspace{-3mm}
\subfigure{\includegraphics[width=0.6\textwidth]{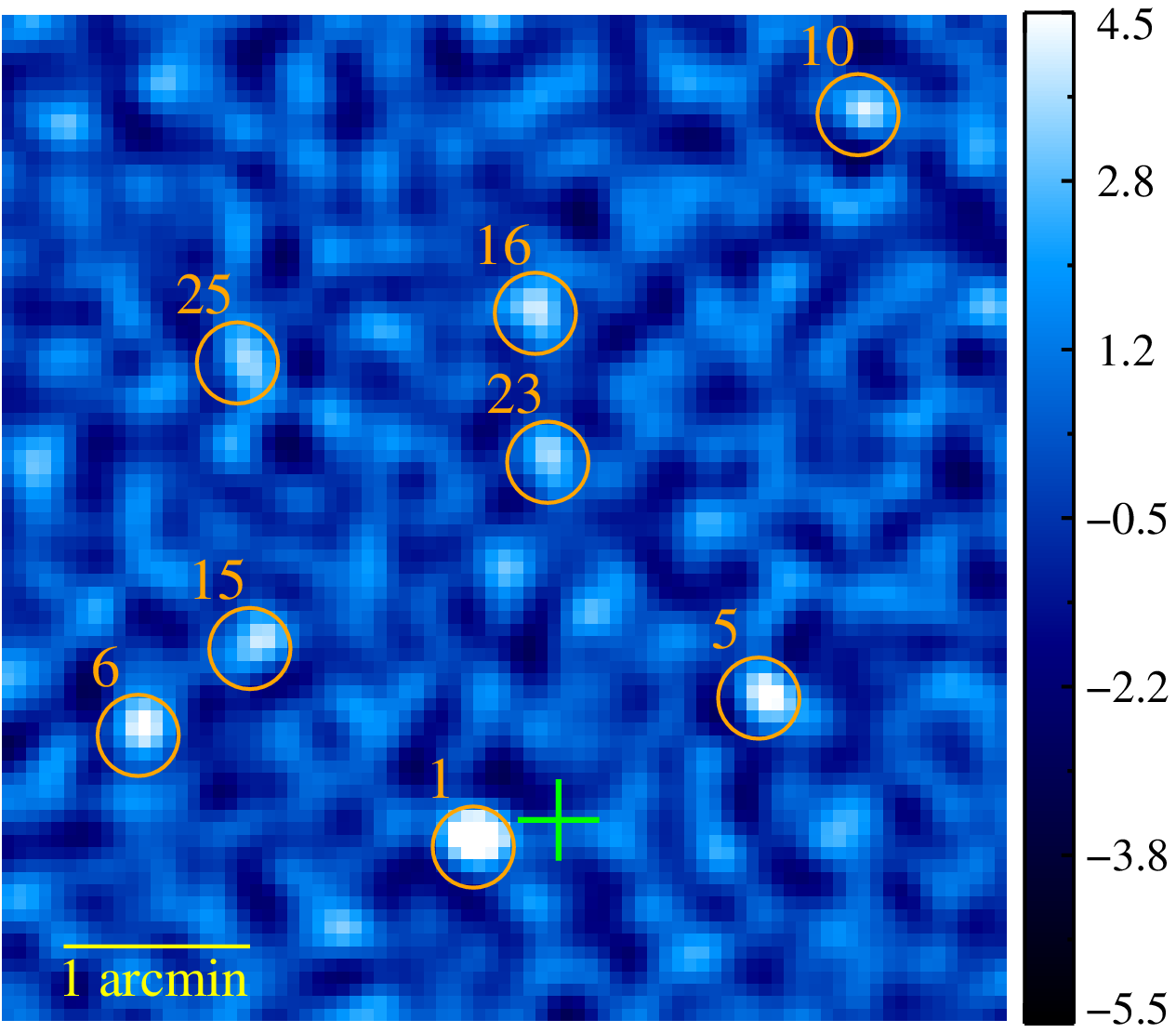}}
\hspace{-125mm} \caption{Left: 850\um\ signal-to-noise ratio map of the
RCS\,2319+00 supercluster overlaid with the red-sequence contours
indicating the cluster positions in red. We highlight all SMGs detected
at $>$3.5$\sigma$ in the 850\um\ map with orange circles. The green contours show
the fractional overdensity of SMGs compared to the average density over the
field, starting at $\delta=-0.3$ and increasing by levels of 0.7. The white circle represents the region from which sources are extracted.
Right: a zoom-in on the overdensity of SMGs surrounding RCS\,2319--C, labeled with their ID in the 850\um\ catalog. Given its proximity to the centre of RCS\,2319--C (the BCG is marked by the green cross), the
brightest source in the catalog, SMM J2319.1, is possibly lensed by the cluster
potential.} \label{fig:s2map}
\end{figure*}

Here, we report early findings from a wide-field SCUBA-2 and {\it
Herschel}-SPIRE survey of a spectacular merging $z=0.9$ supercluster, RCS\,231953+00
(hereafter RCS\,2319+00; \citealp{Faloon13}). The structure
comprises three massive ($\sim$5$\times10^{14}$M$_\odot$), X-ray bright galaxy
clusters \citep{Hicks08}, namely RCS\,231953+0038.1,  RCS\,232003+0033.5, and RCS\,231946+0030.6 (hereafter RCS\,2319--A, RCS\,2319--B, and
RCS\,2319--C). The three clusters are separated by less than 3 projected Mpc and are
expected to merge to form a single Coma-mass cluster by $z\sim0.5$
\citep{Gilbank08}. RCS\,2319--A is a known strong-lensing cluster
\citep{Gladders03}, and the three cores appear to be connected by filamentary
structure and satellite groups. The wealth of data and discoveries surrounding RCS\,2319--A is extensive; RCS\,2319--C, on the other hand, has thus far been comparatively inconspicuous due to a lack of wide-field coverage encompassing the entire structure.

RCS\,2319+00 affords a unique opportunity to study the physics of galaxy
formation across the full range of galaxy environments at $z\approx1$, as well
as more distant galaxies lensed by the cluster potential. In this Letter, we
present a serendipitous discovery of an overdensity of SMGs behind
RCS\,2319--C, as well as a bright SMG close to the cluster core. 
This overdensity could result from a large number of lensed SMGs behind the cluster or represent a possible distant protocluster. This discovery along the line-of-sight to the RCS\,2319
supercluster is a fortunate coincidence: it allows us to investigate the
properties of SMGs within the densest environments in the Universe at two
epochs within the same field, with the benefit that the more distant structure
might be partially lensed by the supercluster. We assume a $\Lambda$CDM cosmology with $(\Omega_m, \Omega_\Lambda, h) = (0.3,0.7,0.7)$.

\section{Observations and data reduction}

\subsection{The RCS\,2319+00 Supercluster}
\label{sec:rcs2319}

Originally discovered in the Red-sequence Cluster Survey (RCS-1;
\citealp{Gladders05}) and presented in \cite{Gilbank08}, the RCS\,2319+00
structure now has extensive follow-up observations. Much of the work
has focused on the northern-most cluster (RCS\,2319--A); it was
revealed to be a remarkable strong-lensing cluster with three gravitationally
lensed radial arcs \citep{Gladders03}, and has a significant weak lensing
signal \citep{Jee11}.  SCUBA imaging of the core of RCS\,2319--A unveiled a
candidate lensed SMG \citep{Noble12} and \textit{Herschel}-SPIRE imaging revealed a
2.5\,Mpc filament of SMGs connecting RCS\,2319--A to its
eastern companion, RCS\,2319--B \citep{Coppin12}.

\subsection{850\um\ SCUBA--2 observations}

SCUBA-2 observations were conducted at the JCMT in Band 2 weather ($0.05<\tau_{\rm
225\,GHz}<0.08$) between 17--21 September 2012 using the 30$'$
PONG mapping pattern. The total mapping time was 7.75\,hr, split into 11$\times$ 40\,min
scans. Individual scans are reduced using the dynamic iterative map-maker
({\it makemap}) of the {\sc smurf} package \citep{Chapin13}
following the procedure outlined in \cite{Geach13}. These scans are co-added
in an optimal, noise-weighted manner, using the {\tt MOSAIC\_JCMT\_IMAGES}
recipe in the {\sc Picard} environment. Finally, to improve the detectability
of faint point sources, we use {\tt SCUBA2\_MATCHED\_FILTER}, which removes large angular scale varying pattern noise in the map by smoothing with a 30$"$ Gaussian kernel, subtracting this, and then
convolving the map with the 850\um\ beam. The average exposure time over the
`nominal' 30$'$ mapping region in the co-added map is $\approx$$10$\,ksec,
reaching a central depth of 1.5\,mJy.

\subsection{{\it Herschel}-SPIRE observations}

The \textit{Herschel}-SPIRE \citep{Griffin10} data were taken on January 2, 2013, with a total of
8.1 hours of integration time over five dithered maps at 250, 350, and
500\um\ (OBSIDs 1342258348, 1342258349, 1342258350, 1342258351, 1342258352). The observations cover 30$'\times$30$'$, including all three cluster
cores, and were carried out in array mode using the nominal scan speed. Each
map is reprocessed individually using {\sc hipe v10.0} \citep{Ott10} and the latest
calibration tree. One map has significant artifacts and requires
a linear polynomial fit during the destriping process; all other maps are
destriped using a zeroth-order fit. All Level 1 scans are merged and final
mosaiced images created using the na\"{\i}ve mapper with default pixel sizes of
6$''$, 10$''$, and 14$''$, at 250, 350, and 500\um, respectively.

Point sources are extracted using SUSSEXtractor \citep{Savage07} at a
relatively low detection threshold of 3.5$\sigma$ to maximize counterpart
completeness. The source list is passed to a timeline fitter that utilizes the
merged Level 1 timeline data to fit a Gaussian at the source position. For
sources below 30\,mJy, the flux is measured as the peak on the image after
smoothing with a convolution kernel and using a sub-pixel correction factor.
This has been found to be a more reliable flux estimate for faint sources
(SPIRE Webinar, private communication). The uncertainties on the fluxes are
estimated from pixel noise added in quadrature with a nominal confusion noise
of 5.8, 6.3, and 6.8\,mJy at 250, 350, and 500\um, respectively
\citep{Nguyen10}.

\begin{figure}
\includegraphics[width=8.7cm]{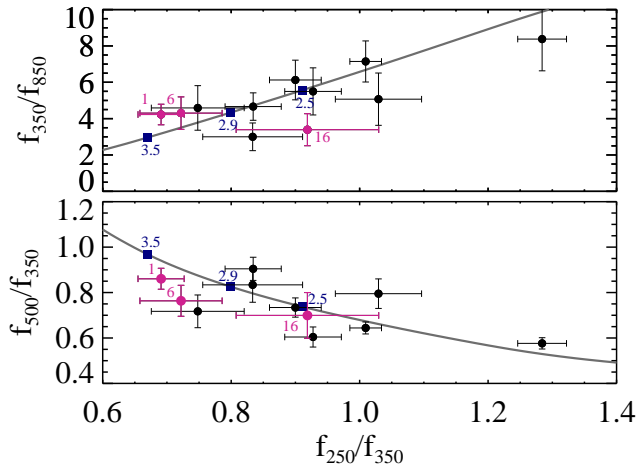}
\caption{Submm/far-IR colour ratios for all SMGs that have
unblended SPIRE counterpart emission in the top panel, and solely SPIRE colors
in the bottom panel. Both plots show the change of these colour ratios (gray
line) with redshift using the median luminosity template from \citet{Chary01},
where the navy squares denote $z=2.5, 2.9, 3.5$. The possible protocluster
members are highlighted in pink, and are at the higher-$z$ end of the
distribution with similar far-IR/submm colours. Confusion errors are not shown
for clarity.} \label{fig:colors} \end{figure}

\begin{figure*} \hspace{-13mm}
\subfigure{\includegraphics[width=0.34\textwidth]{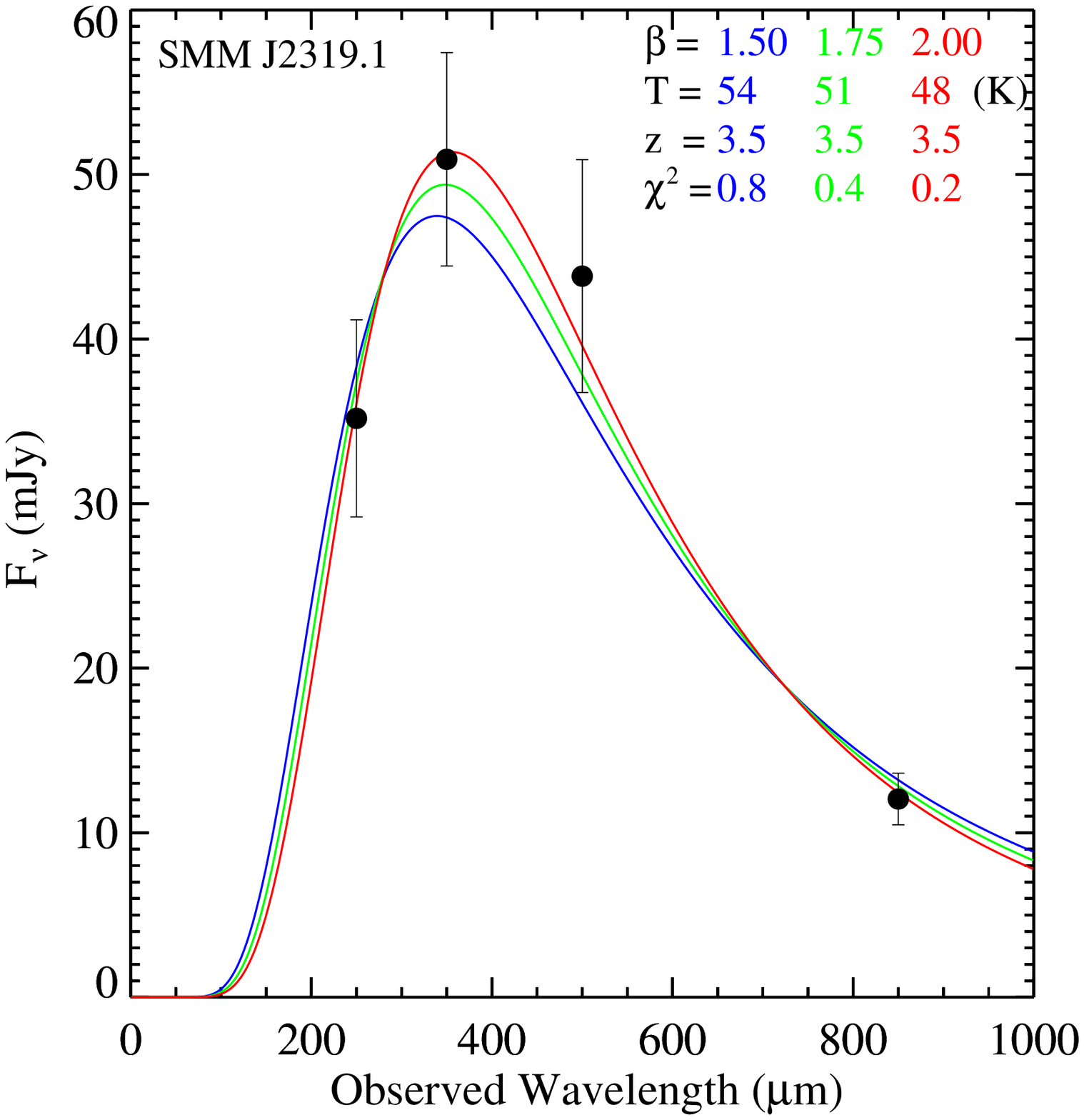}}
\hspace{-12mm}
\subfigure{\includegraphics[width=0.34\textwidth]{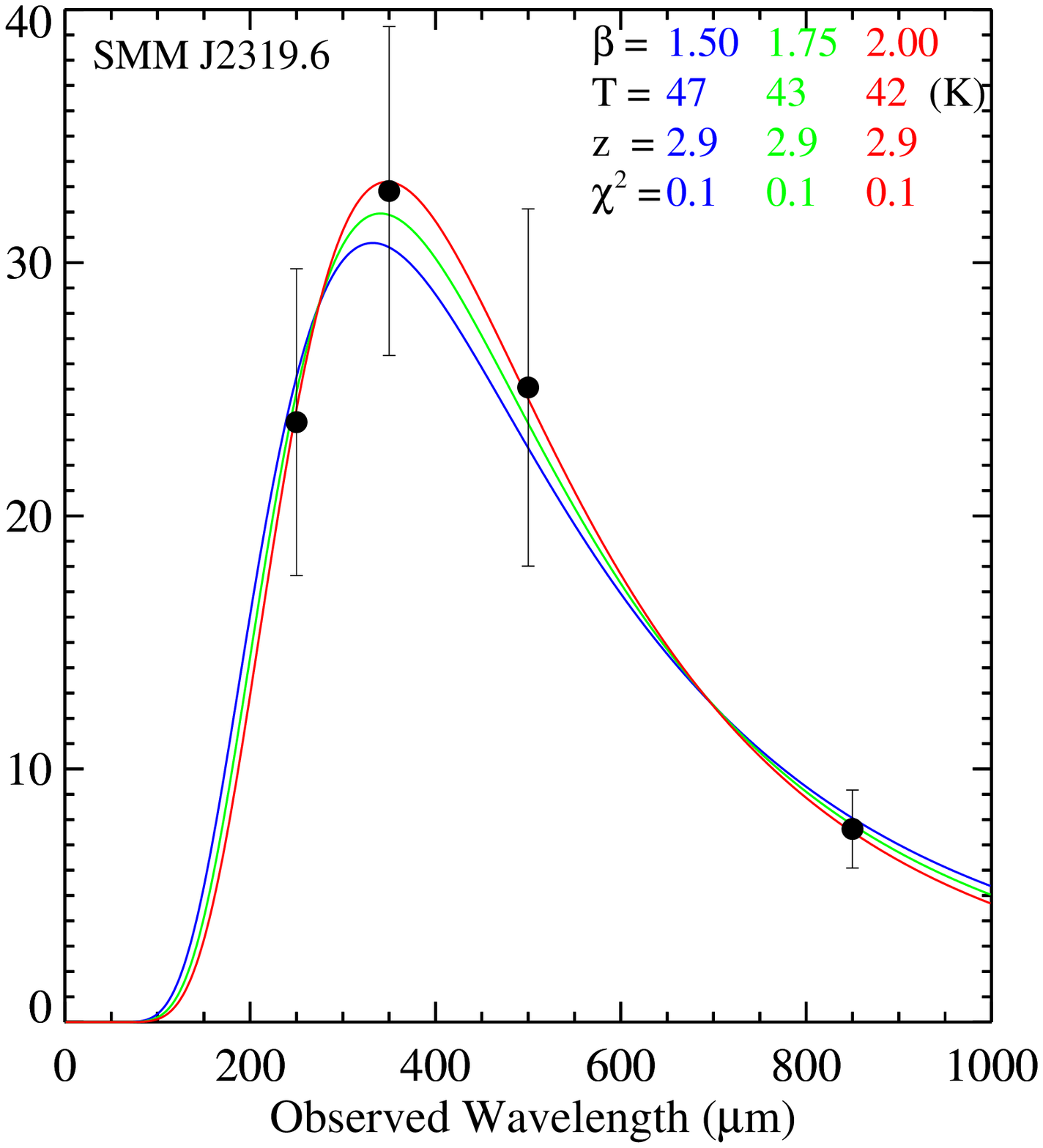}
}\hspace{-12mm}
\subfigure{\includegraphics[width=0.34\textwidth]{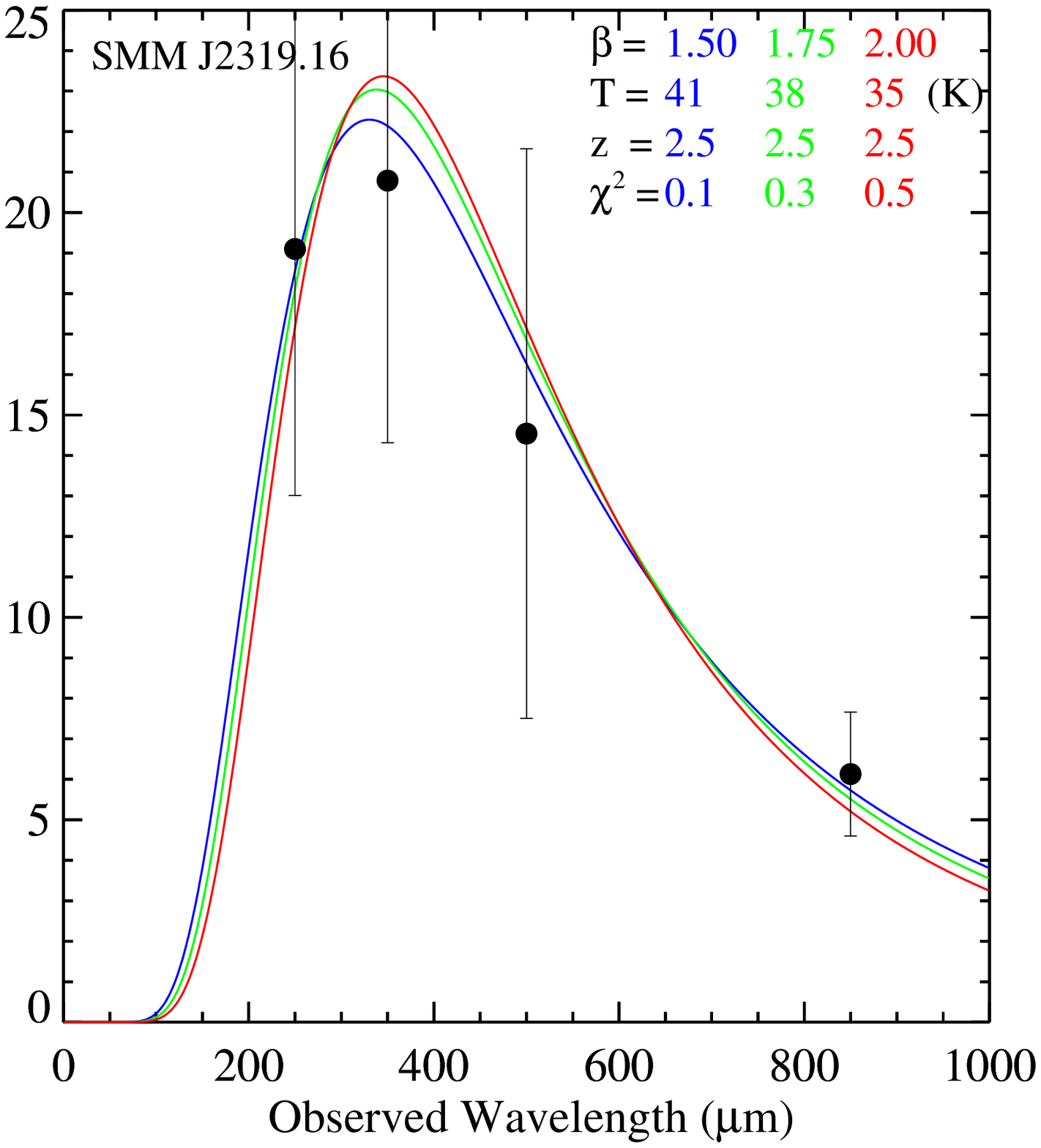}
}\hspace{-11mm} \caption{Modified blackbody fits to the SPIRE and SCUBA-2
fluxes for the three possible protocluster members, using the technique
described in Section~\ref{sec:proto}. The parameters for the best fit values
with $\beta$ fixed at 1.5 (blue), 1.75 (green) and 2.0 (red) are listed in
each panel. The lowest reduced $\chi^2$ values are given by $z=[3.5, 2.9, 2.5]
\pm0.6$ and $T_{\rm d}=[48, 43, 41] \pm 8$ K for SMMs J2319.1, 2319.6, and
2319.16, respectively. We note that SMM J2319.1 has only a weak radio
detection, just below the catalogue limit; we therefore assign it the
3$\sigma$ rms limit of 45$\mu$Jy.} \label{fig:seds} \end{figure*}

\section{Analysis and results} 

Point sources are extracted from the central $\sim$25$'$ of the beam-convolved
map, where the sensitivity is fairly uniform and the noise is $<3$$\times$ the
central $1\sigma$ r.m.s., corresponding to a total uniform area of
$\sim$473\,arcmin$^2$. We detect 29 point sources at 850\um\ at a
significance of $>$3.5$\sigma$, 16 of which are at $>4$$\sigma$. The
detections are indicated by orange circles in Figure~\ref{fig:s2map} and named
in order of descending signal-to-noise ratio.  We quantify the false-detection rate by running the detection algorithm on jack-knife versions of the map (details given in \citealp{Geach13}).  Within the source detection area, we find a false-detection rate of 4.5\% (20\%) at 4.0$\sigma$ (3.5$\sigma$).  We note that both these rates are 25\% lower where the noise is $<2$$\times$ the central r.m.s., encompassing all but one of the SMGs.

\subsection{SCUBA--2/SPIRE source identification}

Very Large Array radio imaging at 1.4\,GHz covers the entirety of the SCUBA-2
and SPIRE maps, although with roughly 2$\times$ the noise at the edges due to
primary beam attenuation \citep{Noble12}. We also exploit deep IR imaging from the Multiband Imaging Photometer (MIPS) aboard \textit{Spitzer}
\citep{Webb13}. The extensive, multi-wavelength counterpart identification process, including the completeness and false-detection rate,
will be presented in Noble et al.\ (in preparation); we provide only a brief
description in this communication.

Given the $\approx$15$''$ beam at 850\um, we search for radio and 24\um\
emission within 10$''$ of the SCUBA-2 positions, ensuring we detect all
possible counterparts \citep{Ivison07, Biggs11}.  While a 10$''$ search radius is generous, in
practice we find that all the counterparts to $>$4$\sigma$ SMGs are within 6$''$, with an average offset of 3$''$. We robustly detect a 24\um\
and/or radio counterpart for 20 of the 29 (70\%) sources within our catalog.
Of the remaining SCUBA-2 detections, eight lack MIPS coverage (and have no
radio detection), and one source eludes any counterpart emission. For cases where multiple 24\um\ sources are detected within the
search area (and no radio emission), we assign the nearest source as the most
likely counterpart, which in all occurrences is also the brightest 24\um\
detection within the search area. 

Counterpart SPIRE emission can further verify the validity of each SCUBA-2
source, although it is not a requirement as 20--60\% of $z>2$ SMGs are
undetectable with SPIRE \citep{Casey12}. Given the large SPIRE beam, we search within the entire SCUBA-2 beam for 250--500\um\
emission, resulting in detections in one or more SPIRE bands for 18 of the 29
850\um\ sources. Four of the SMGs lacking MIPS coverage or a radio
counterpart are detected with SPIRE, yielding a final catalog of 24 (80\%)
SCUBA-2 sources with counterparts in another band. Any emission that is
blended or confused in SPIRE (four cases in total) is omitted from the
counterpart catalog for the purposes of far-IR SED fitting (see
\S\ref{sec:proto}).

\subsection{Evidence of a line-of-sight, submillimetre-bright protocluster}
\label{sec:proto}

The average surface density of 850\um\ sources across the 30$'$ RCS\,2319+00
field is consistent with that expected from the number counts measured in
blank-field submm surveys \citep{Coppin06}; however, it is clear that there are large variations
in the local surface density on scales of several arcminutes. Indeed, there
appears to be a local relative overdensity of 850\um\ detections in the
vicinity of RCS\,2319--C. To quantify this, we create a Gaussian smoothed
($\theta=4'$) surface density map, normalized to the fractional overdensity of
SMGs: $\delta_{\rm peak} = (\rho-\bar{\rho})/\bar{\rho}$ (Figure~\ref{fig:s2map}).

There is a $\delta_{\rm peak}=3.5$ close to RCS\,2319--C. 
To assess the significance of this local peak, we generate
simulated catalogues of the same size as the real map, and with the same source density as the detected SMGs.  In addition to the shot noise properties of the sources, we also include the effects of clustering on linear scales, assuming a redshift distribution and bias factor matching the current best estimate
for 850\um-selected SMGs \citep{Hickox12}.  We check the clustering properties of the fake catalogues by applying the Landy-Szalay estimator \citep{Landy93} and find the angular correlation function $w({\theta})$ to be consistent with \cite{Hickox12} at $\theta\gtrsim1'$, matching the scale of the candidate protocluster.  We generate 1000 catalogue realisations, each time evaluating the
convolved surface density estimate, as described above. The
total number of density peaks found with $\delta\geq\delta_{\rm peak}$ is an
estimate of the significance of finding a local overdensity of magnitude
3.5 or greater in our map. We find nine such peaks over all 1000 realisations, meaning the SMG overdensity is very unlikely to be due to Poisson noise or clustering from large scale structure, significant at a level of 99\%. In other words, such a structure is rare, even
given the known correlation function of SMGs, and thus could point to the
presence of a real physical association of SMGs tracing a highly biased
structure.

With four submm bands, we can also compare the submm colours of the SMGs
around RCS\,2319--C to test if they have similar far-IR SEDs (and therefore
possibly similar redshifts) in addition to a spatial correlation. In
Figure~\ref{fig:colors}, we plot the SPIRE and SCUBA-2 colours of 11 SMGs that
are detected in all three SPIRE bands.  We do not include 8 sources that have blended SPIRE emission. The sources within the
overdensity are highlighted in pink, and lie at the higher redshift end of the
colour distribution compared to the
other SMGs in the field. SMM\,J2319.1 and SMM\,J2319.6 are quite distinct,
while SMM\,J2319.16 tends towards the bluer end of the $S_{250}/S_{350}$
distribution, but is still within $\sim$1$\sigma$ of the other putative
protocluster galaxies.

We estimate the redshifts of these sources by fitting to the submm photometry,
assuming the SED can be modeled by a single temperature modified blackbody,
with a smooth transition into a power law on the Wien side. 
Because of the degeneracy in
the $z$--$T$--$\beta$ parameter space, we perform three fits, each time only
allowing $z$ and $T$ to be free parameters ($20<T<60$\,K), while fixing
$\beta$ at 1.5, 1.75 and 2, spanning the likely range of real galaxies
\citep{Dunne00}. We further improve upon the estimate by
putting prior constraints on $z$ and $T$ from independent measurements, as
shown in \cite{Roseboom12}. We estimate a submm-radio photo-$z$ using the
spectral index relation from \cite{Carilli99}, assuming $\alpha_{\rm 1.4GHz} =
-0.8$, $\alpha_{850\mu m} = \beta+2$, and a typical uncertainty of $\Delta z
=1.0$ \citep{Aretxaga07}. We also simultaneously measure the total
(8--1000\um\ rest-frame) IR luminosity to estimate $S_{60}/S_{100}$
using the empirical relation in \cite{Roseboom12} from a SPIRE-selected
sample. We then convert the flux ratio into a dust temperature from the
modified blackbody described above and conservatively assume a 10\,K scatter
in $T_d$. 

This fitting algorithm greatly reduces the degeneracy between $z$ and $T_d$
and produces typical uncertainties of $\Delta z=0.6$ and $\Delta T_{d} =
8$\,K. The best fit SEDs are shown in Figure~\ref{fig:seds}. The lowest values
of $\chi^2/\nu$ yield $z=$[3.5, 2.9, 2.5] and $T_d=$[48, 43, 41] K for
SMMs\,J2319.1, J2319.6, and 2319.16, respectively. These redshifts
are all within $\sim1\sigma$ and further support the existence of an
SMG protocluster behind RCS2319-C at $z>2.5$. Although this
redshift range is representative of the typical SMG \citep{Chapman05}, we
emphasize the similarity of submm colours for these SMGs, which isolates them
from the rest of the sample through a consistent comparison.  The SED fit also provides an estimate of the IR luminosity which can be converted to a SFR (Salpeter IMF; \citealp{Kennicutt98}); we find all three candidate protocluster members are highly active, with SFRs of 2500, 1100, and 500\,\myr (assuming no AGN contamination).  We note that these could be over estimated if any of the SMGs are strongly lensed.

\subsection{A submillimetre-bright, strongly lensed galaxy}
\label{sec:arc}

Within this putative protocluster lies the brightest 850\um\ source in the
catalogue, SMM\,J2319.1, with $S_{850}=12.05\pm1.56$\,mJy. This source is only
28$''$ away from the X-ray peak of RCS2319--C and could be strongly lensed by the cluster potential, though it is only
1.5$\times$ as bright as the second brightest protocluster member.
SMM\,J2319.1 is coincident with an extremely red spur-like feature---possibly the optical/near-IR counterpart to the SMG. It is considerably redder than the red-sequence galaxies in the cluster, implying
that it is probably at higher redshift. Indeed, the SED fit places this source
at $z=3.5$, although (sub)mm spectroscopic identification will be required to
accurately determine the redshift.

\begin{figure} \centering
\includegraphics[width=0.47\textwidth]{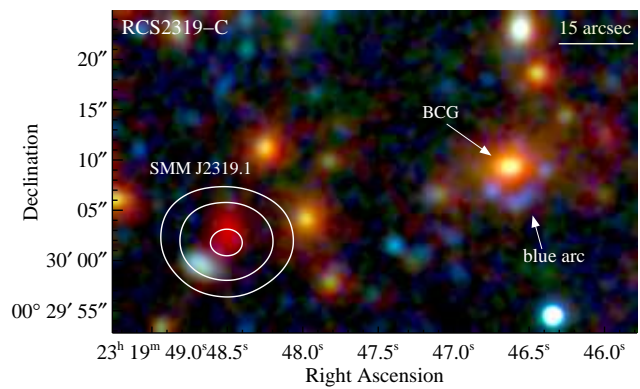} \caption{A
{\it gK}[4.5] colour composite (from CFHT and IRAC) of the centre of RCS\,2319--C, showing the
position of with SMM J2319.1, the possible lensed SMG (850\um\ S/N contours
starting at $5\sigma$, increasing by levels of $1\sigma$), Note the very red
optical/near-IR emission coincident with the SCUBA-2 detection. We highlight
the BCG, with a blue arc-like feature to the south, further indicating that
this could be a strong-lensing cluster.} \label{fig:arc} \end{figure}

In Figure~\ref{fig:arc} we present a {\it gK}[4.5] composite image of the
cluster core.  850\um\ S/N contours mark the position of the galaxy.  
There are hints of a blue arc-like feature around the BCG, indicating that this may be a strong lensing cluster. We lack an accurate mass model for this cluster
(high-resolution optical imaging only exists for RCS\,2319--A), but assuming
an isothermal sphere with $\sigma_v=759$\,km\,s$^{-1}$ \citep{Faloon13} and
$z=3.0$ for the source plane, we calculate an Einstein radius of $\sim9''$,
which is consistent with the distance between the BCG and blue arc. We emphasize
that the projected overdensity of SMGs around RCS2319--C could also be due to
boosted counts from lensing of the random field, rather than a physical
protocluster (spectroscopic confirmation will answer this question).
However, the peak of the overdensity is offset from the core of RCS\,2319--C
by $>$1$'$, while the average lensing magnification
beyond the central 30$''$ is expected to be $<2$ for similarly massive
clusters \citep{Noble12}.

\section{Summary}

This Letter presents the first wide-field SCUBA-2 850\um\ map of a high-$z$
galaxy cluster field. We target a rare, three-component supercluster at
$z=0.9$ -- RCS\,2319+00 -- detecting 29 SMGs, the majority of which have
robust counterparts at 24$\mu$m, 1.4\,GHz and/or in the {\it Herschel}-SPIRE
bands. Previous work on RCS\,2319+00 has focused on the northern core,
RCS\,2319--A, which is a well-known strong-lensing system, whereas the new
SCUBA--2 map reveals that RCS2319--C, the southern-most component, also has some
interesting traits. We make two discoveries:

\begin{enumerate}

\item RCS2319--C has features indicative of a strongly lensing cluster,
with a distinct blue arc just below the BCG (at a radius
consistent with the Einstein radius expected for this cluster). We report the
discovery of a bright ($S_{850}\approx 12$\,mJy) SMG 28$''$ from the core, associated
with a very red optical/near-IR counterpart that is likely to be a lensed
galaxy at $z\approx3$. Thus, it offers the rare
opportunity to study the properties of the SMG in much finer detail than would
otherwise be possible.

\item There is a significant local overdensity of SMGs in the vicinity of RCS2319--C, with a peak of $\delta=3.5$ in density contrast when smoothed at 4$'$.
Simulations indicate that there is a $<$1\% chance of finding a similar
structure in a (clustered) blank field of the same area. We estimate the redshifts
for three of the sources within this overdensity by fitting the observed
250--850\um\ photometry with a modified blackbody SED, finding them consistent with $2.5<z<3.5$. They have high IR luminosities, corresponding to SFRs ranging from $500-2500$\,\myr. We speculate that the SMGs are part of a
physical association at $z\approx3$, perhaps signposting a starbursting
protocluster along the line of sight to RCS\,2319--C. This scenario is supported
by recent  clustering measurements which predict the formation of SMGs in compact
protoclusters \citep{Maddox10}. Indeed, SMGs have been found to trace the
underlying distribution of Lyman-$\alpha$ emitters in a $z\approx3.1$
protocluster \citep{Tamura09} and in some cases are
physically associated with Ly$\alpha$ Blobs in these enviroments
\citep{Chapman01,Geach05}.

\end{enumerate}

We thank many people for assistance with the SCUBA-2/SPIRE data, including the JAC staff and telescope operators, and the SPIRE ICC, especially David Shupe.  We also thank Gaelen Marsden for useful discussions. JG acknowledges support from a Banting Fellowship administered by NSERC. TW is supported by the NSERC Discovery Grant and the FQRNT
Nouveaux Chercheurs program. KC receives support from the endowment of
the Lorne Trottier Chair in Astrophysics and Cosmology at McGill and NSERC.

\bibliography{mnemonic,references}
\bibliographystyle{apj}

\label{lastpage} \end{document}